\documentclass[conference]{IEEEtran}
\IEEEoverridecommandlockouts
\usepackage{cite}
\usepackage{amsmath,amssymb,amsfonts}
\usepackage{algorithmic}
\usepackage{graphicx}
\usepackage{textcomp}
\usepackage{xcolor}
	
\usepackage[hidelinks]{hyperref}

\usepackage[]{todonotes} 
\usepackage{acronym}
\usepackage{graphicx}
\usepackage{multirow}
\usepackage{colortbl}
\usepackage{nameref}
\usepackage{subfig}
\usepackage{amsmath}
\usepackage{breqn}
\usepackage{url}
\usepackage{hhline}
\usepackage{tcolorbox}

\def\BibTeX{{\rm B\kern-.05em{\sc i\kern-.025em b}\kern-.08em
    T\kern-.1667em\lower.7ex\hbox{E}\kern-.125emX}}
\begin{document}

\title{Network Calculus Results for TSN: An Introduction}

\author{\IEEEauthorblockN{Lisa Maile}
\IEEEauthorblockA{\textit{Computer Science 7} \\
\textit{Friedrich-Alexander University}\\
\textit{Erlangen-Nürnberg}, Germany \\
lisa.maile@fau.de}
\and	
\IEEEauthorblockN{Kai-Steffen Hielscher}
\IEEEauthorblockA{\textit{Computer Science 7} \\
\textit{Friedrich-Alexander University}\\
\textit{Erlangen-Nürnberg}, Germany \\
kai-steffen.hielscher@fau.de}
\and
\IEEEauthorblockN{Reinhard German}
\IEEEauthorblockA{\textit{Computer Science 7} \\
\textit{Friedrich-Alexander University}\\
\textit{Erlangen-Nürnberg}, Germany \\
reinhard.german@fau.de}
}

\maketitle

\begin{abstract}
Time-Sensitive Networking (TSN) is a set of standards that enables the industry to provide real-time guarantees for time-critical communications with Ethernet hardware. TSN supports various queuing and scheduling mechanisms and allows the integration of multiple traffic types in a single network. Network Calculus (NC) can be used to calculate upper bounds for latencies and buffer sizes within these networks, for example, for safety or real-time traffic. We explain the relevance of NC for TSN-based computer communications and potential areas of application. Different NC analysis approaches have been published to examine different parts of TSN and this paper provides a survey of these publications and presents their main results, dependencies, and differences. We present a consistent presentation of the most important results and suggest an improvement to model the output of sending end-devices. To ease access to the current research status, we introduce a common notation to show how all results depend on each other and also identify common assumptions. Thus, we offer a comprehensive overview of NC for industrial networks and identify possible areas for future work.
\end{abstract}


\begin{table}[b]
	\vspace{-4mm}
	\begin{minipage}{0.95\linewidth}
		\begin{tcolorbox}[colback=blue!5!white, colframe=blue!40!black, boxrule=0.5pt]
			\scriptsize
		\textbf{Copyright~\copyright~2022 IEEE}\\ 
Personal use of this material is permitted. Permission from IEEE must be obtained for all other uses, in any current or future media, including reprinting/republishing this material for advertising or promotional purposes, creating new collective works, for resale or redistribution to servers or lists, or reuse of any copyrighted component of this work in other works.\\
\textbf{Citation}:
L. Maile, K. -S. Hielscher and R. German, "Network Calculus Results for TSN: An Introduction," 2020 Information Communication Technologies Conference (ICTC), Nanjing, China, 2020, pp. 131-140, doi: 10.1109/ICTC49638.2020.9123308.
\\
\textbf{Published version:} \url{https://doi.org/10.1109/ICTC49638.2020.9123308}\end{tcolorbox}
	\end{minipage}
\end{table}

\section{Introduction}
\label{sec:intro}
Ethernet plays an increasingly important role in industrial communications as it offers higher data rates, lower costs, and easier integration with existing IT systems than traditional fieldbuses~\cite{bruckner_introduction_2019}. 
However, traditional Ethernet is not designed to offer hard real-time guarantees, which become increasingly important for avionics~\cite{avionics}, automotive~\cite{autotrends}, and train industries~\cite{train} as well as in industrial automation~\cite{iec60802}.
 
With Time-Sensitive Networking (TSN), these distributed time-critical applications can be built based on standard Ethernet technology while allowing multiple traffic classes on the same network ranging from real-time to best-effort. TSN is developed by the 802.1 Working Group and provides additional standards for Ethernet to limit end-to-end delays and jitter by introducing various queuing and scheduling mechanisms. 

To the best of our knowledge, we are the first to generally present the relevance of Network Calculus (NC) for TSN and provide transparency and structure to the multitude of publications in this context. We provide the first survey about literature that derive NC models for TSN schedulers to obtain worst-case (WC) guarantees. Therefore, we structurally explain the results for the two most important NC elements - arrival and service curves-, illustrate their basic application, and summarize the existing papers. Besides, we have improved the arrival curve for aperiodic TSN traffic. 

Be aware that we do not compare the results, e.g. by simulation, but that we emphasize which are the most advanced results and which schedulers have been examined. With the large number of publications, it is not trivial to identify whether
and in which part the publications have different results or improve each other. With this survey, the modeling details of each publication are clarified and the latest results are highlighted and can be used as first references.

To sum it up, our key contributions are:
\begin{enumerate}
\item Discuss the practical relevance of NC for industrial communications with TSN in general.
\item Present a survey about existing literature for NC delays of TSN schedulers.
\item Explain the main NC results and their interdependencies.
\end{enumerate}

Some presentations with published slides~\cite{boudec_application_nodate, zhang_presentation} exist which present rules and mathematical results of NC that are valuable to derive NC bounds in TSN. They mainly focus on the results for a single shaper, Asynchronous Traffic Shaping, and are a nice supplement to this introduction. We extend these presentations with the first introduction to all NC results for TSN and a general discussion about when NC is necessary and how it can be used in TSN. 

The principles of TSN and NC are introduced in this section. Section~\ref{sec:reasons} describes the reasons for NC in TSN and possible applications. Afterwards, Section~\ref{sec:existing_work} defines and explains the details of the derived NC results for TSN, how they depend on each other, and how they can be applied. Thereby, the schedulers and NC definitions are always introduced in the corresponding subsections of Section~\ref{sec:existing_work}. Lastly, Section~\ref{sec:summary} concludes the paper with a summary. 

\subsection{Principles of TSN}
\label{sec:tsn_intro}
The Time-Sensitive Networking Task Group~\cite{tsnWebsite} of IEEE 802.1 was formed to extend the Audio Video Bridging~\cite{avbWebsite} standards and develop new mechanisms for general real-time communications~\cite{teener_heterogeneous_2013}. For an extensive overview of all TSN schedulers, see~\cite{nasrallah_ultra-low_nodate}.

\begin{figure}[!t]
\centering
\includegraphics[trim={0cm 4cm 0cm 0cm} ,clip, page=1, width=2.4in]{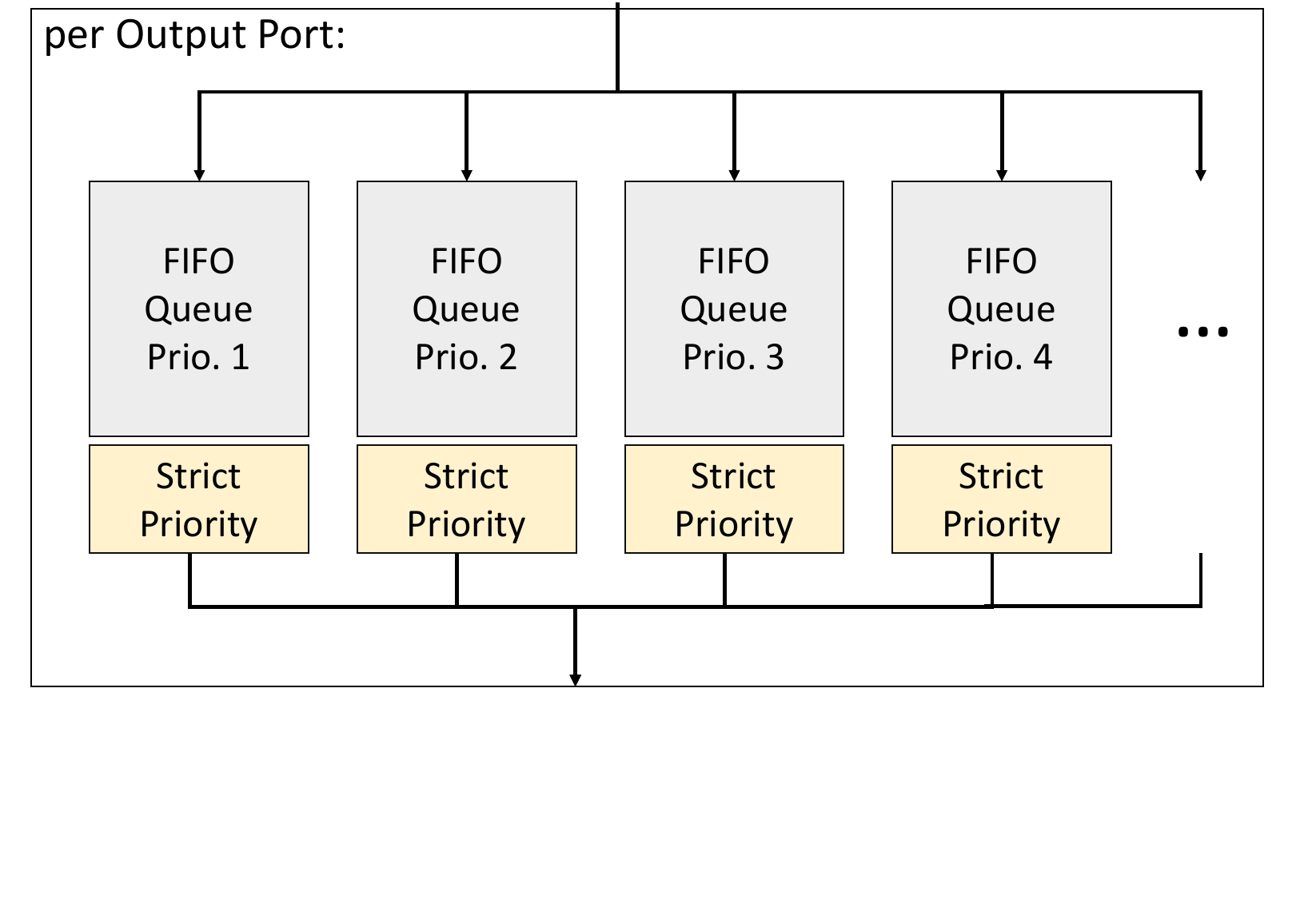}%

\caption{Basic example setup of a TSN scheduler which can be extended with further scheduling mechanisms.}
\label{Fig:TSN}
\end{figure}

In any case, TSN schedulers consist of several FIFO output queues~\cite{802Q}. Fig.~\ref{Fig:TSN} shows the basic structure for scheduling in a TSN output port. Ideally, eight queues are available, one for each priority. Queues always process one traffic class, but if there are less than eight queues, the standard allows multiple priorities to be assigned to one class~\cite{802Q}. The traffic with the highest priority (denoted in this paper as 1) will be transmitted with Strict Priority selection (SP)~\cite{802Q} if all other implemented schedulers have selected the frame as transmission-ready. In real applications, the scheduler in Fig.~\ref{Fig:TSN} has more advanced mechanisms which are introduced in Section~\ref{sec:existing_work}.

\subsection{Introduction to Network Calculus}
\label{sec:nc_intro}

Network Calculus (NC) is a system theory for communication networks. Guaranteed upper bounds can be calculated by modeling flows and nodes as functions. 
NC analyzes network elements such as individual queues, complete nodes, or entire networks. We call the analyzed network element "system" in the following.

Generally, we abstract the incoming data and the scheduling of each system with curves and derive bounds per system. The real incoming and outgoing traffic for a system is referred to as $R(t)$ and $R'(t)$, which are cumulative functions that model the number of bits during the time interval [0, t].
The real data flow is bounded with maximum arrival and minimum service curves. This enables the derivation of safe upper bounds for WC end-to-end delays and queuing backlogs. 

\begin{figure}%
    \centering
    \subfloat[Convolution $x \otimes y$.]{{\label{fig:conv}\includegraphics[width=1.55in, page=2]{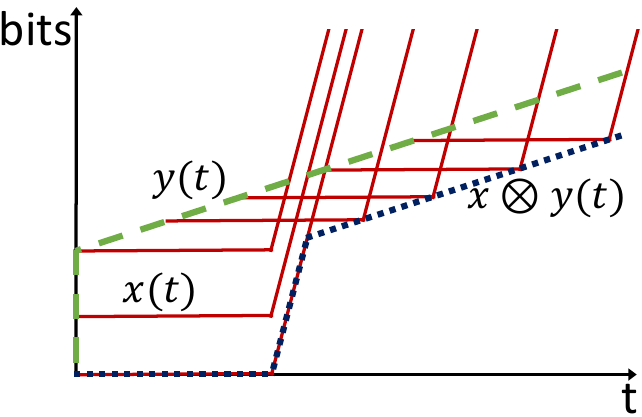} }}%
    \qquad    
        \subfloat[Deconvolution $x \oslash y$.]{{\label{fig:deconv}\includegraphics[width=1.55in, page=3]{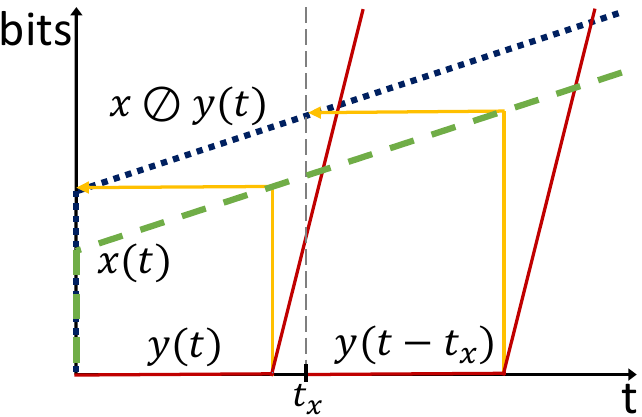} }}%
    \label{fig}%
    \caption{Basic NC operations.}
\end{figure}

NC is often named together with the min-plus algebra. Min-plus algebra is a commutative half-ring. It defines mathematical operators differently. For example, the convolution~(1) and deconvolution~(2) are defined as
\cite{boudec_network_2012, borys_principles_2011, boudec_made_easy, boudec_I, BOUILLARD2010306}:
\begin {equation} (x \otimes y)(t) := \inf_{0\le s \le t} \{x(t-s) + y(s)\} \end {equation} 
\begin {equation}\label{equ:deconv} (x \oslash y)(t) := \sup_{0\le s}\{x(t+s) - y(s)\} \end {equation} 
They are commonly used in NC to shorten definitions. Illustratively, these functions mean the following. The convolution is commutative, therefore, $(x\ \otimes\ y)(t)$ places one of the two functions at each point of the other function and the resulting curve is then the minimum~\cite{kellerer_network_2016}, as illustrated in Fig.~\ref{fig:conv}.
The deconvolution $(x\ \oslash\ y)(t)$ increases the function $x$ at every point $t_x \in [ 0,t ]$ by the maximum vertical distance that can occur between $x$ and $y$ with $y$ shifted by $t_x$, see Fig.~\ref{fig:deconv}.


\section{Reasons for NC in TSN}
\label{sec:reasons}

This section explains how NC can be applied, with each application highlighted in \textit{italics}.
Generally, we categorize schedulers into deterministic and non-deterministic. Be aware that in this case, non-determinism does not mean that we have random or unknown behavior of the schedulers, but that the scheduling for a flow changes due to the interference of other flows. With NC, \textit{guaranteed upper limits for delays can be derived even if the scheduling behaves non-deterministic}.

With the variety of schedulers that are possible in TSN, their effects on maximum delays need to be investigated. The combination of different TSN scheduling mechanisms leads to non-trivial complexity in the sense of WC analysis, for which NC can be used as a formal framework. With NC, we see how scheduling mechanisms and parameters influence the delay~\cite{reviewer_journal}. Thus, NC analysis can \textit{support network designers with information to choose the fitting scheduler mechanisms and queue mappings} for the required scenarios.

In contrast to simulation or heuristic methods, NC can derive \textit{reliable and replicable evidence of upper limits} and correct real-time behavior for critical applications. Since we model safe upper and lower bounds of incoming and outgoing traffic, the delay and buffer sizes are reliably maximized even though they might be too pessimistic.

To enable real-time communications, the forwarding devices are configured with dedicated scheduling mechanisms for which the fitting configuration parameters need to be found. NC can perform the part of \textit{testing the validity of a configuration}, meaning it checks whether the current parameters can meet all latency guarantees without needing simulations or prototypes. Thus, NC can be used to support and validate the configuration process in TSN networks. 

Current configuration approaches for time-aware scheduling require a strict timely separation of flows to keep the delay calculation simple~\cite{craciunas_scheduling_2016}. With enhanced delay analysis as NC, the \textit{use of more complex configurations} is possible. As we will see in Section~\ref{sec:existing_work}, we can combine scheduler mechanisms and allow less strict separation for the flows and, thus, \textit{increases the solution space for time-triggered schedulers}.

As also explained later, NC can derive guaranteed maximum buffer sizes and thus allow for correct scaling of networks. Thereby, it can \textit{eliminate packet loss due to buffer overflows}.

NC calculations can be implemented in the centralized or hybrid configuration models, in particular, in the so-called \textit{Central Network Configurator} (CNC)~\cite{CNC_TSN}. This component has a complete overview of the network and is used to reserve flows in the bridges. \textit{In the CNC, an automatic configuration of the network could be supported by NC}.


Besides, TSN uses the stream reservation protocols MSRP~\cite{Qat, Qcc}/MSRP++~\cite{MSRPplus} and RAP~\cite{Qdd} to reserve flows decentrally without CNC. Therefore, these protocols include a field that is updated at each hop that denotes the maximum accumulated latency. It is not yet defined how this field is to be filled by the forwarding devices. To guarantee that all delays are included, the results of NC could potentially be used to \textit{update the maximum accumulated latency field of stream reservation protocols in TSN}.

\section{Existing NC Work for TSN}
\label{sec:existing_work}
Many results have already been derived with NC for TSN, but since existing publications have been continuously improved or revised, it is necessary to determine the current state of the art. We have surveyed the scope and results of all relevant publications and explain their main modeling approaches.

Every publication uses own notations for their results. We unify the notations and variables and rearrange the formulas to make understanding as easy as possible. Thereby, this section explains each result with illustrative explanations of the basic ideas. We also present formulas in a way that shows the interdependencies of existing literature and how the single curves in each work were derived from each other. Due to their relevance, we focus our explanation on arrival and service curves that we will also define in the corresponding subsections. We further propose an improvement to the arrival curve for aperiodic traffic and show a small example how NC can be applied. Afterwards, we shortly mention which additional improvements or curves, besides arrival and service, are offered in each paper in Section~\ref{sec:details} where we also provide a table for quick reference.

The general assumptions of all publications which derive NC results for TSN are the following: The service curves are modeled to define the behavior of one output queue with incoming traffic from potentially multiple input ports. The analyzed system is, therefore, one output queue for which arrival and service were defined. Thereby, it is assumed that each queue handles one priority. Due to the aggregated scheduling of flows with the same priority in one queue, only class-based service curves which are the same for all flows of one priority are derived. Besides, it is assumed that the parameters for the curves are known and all links have the same bandwidth.
\subsection{Arrival Curve for TSN}
\label{sec:arrival}
\begin{figure*}%
    \centering
    \subfloat[WC sending behavior of TSN flow.]{{\label{sending}\includegraphics[width=5cm,]{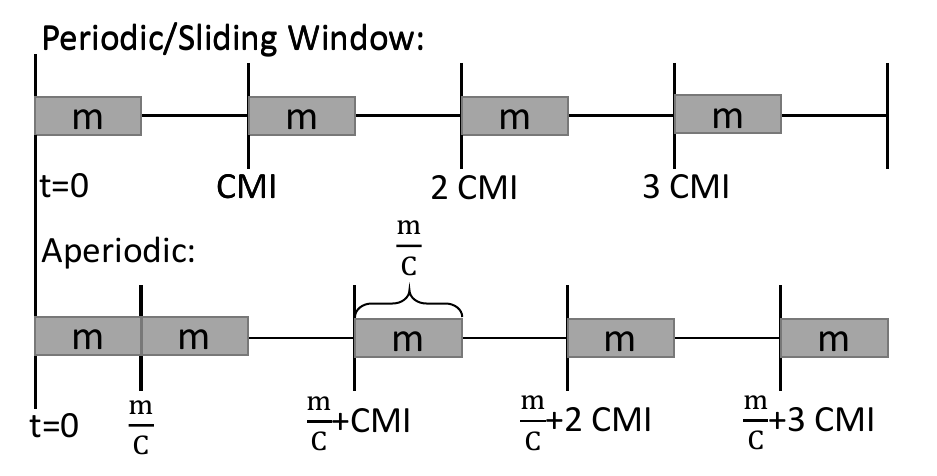} }}%
    \qquad
    \subfloat[Arrival curves for periodic traffic.]{{\label{per_arr}\includegraphics[width=4.9cm, page=1]{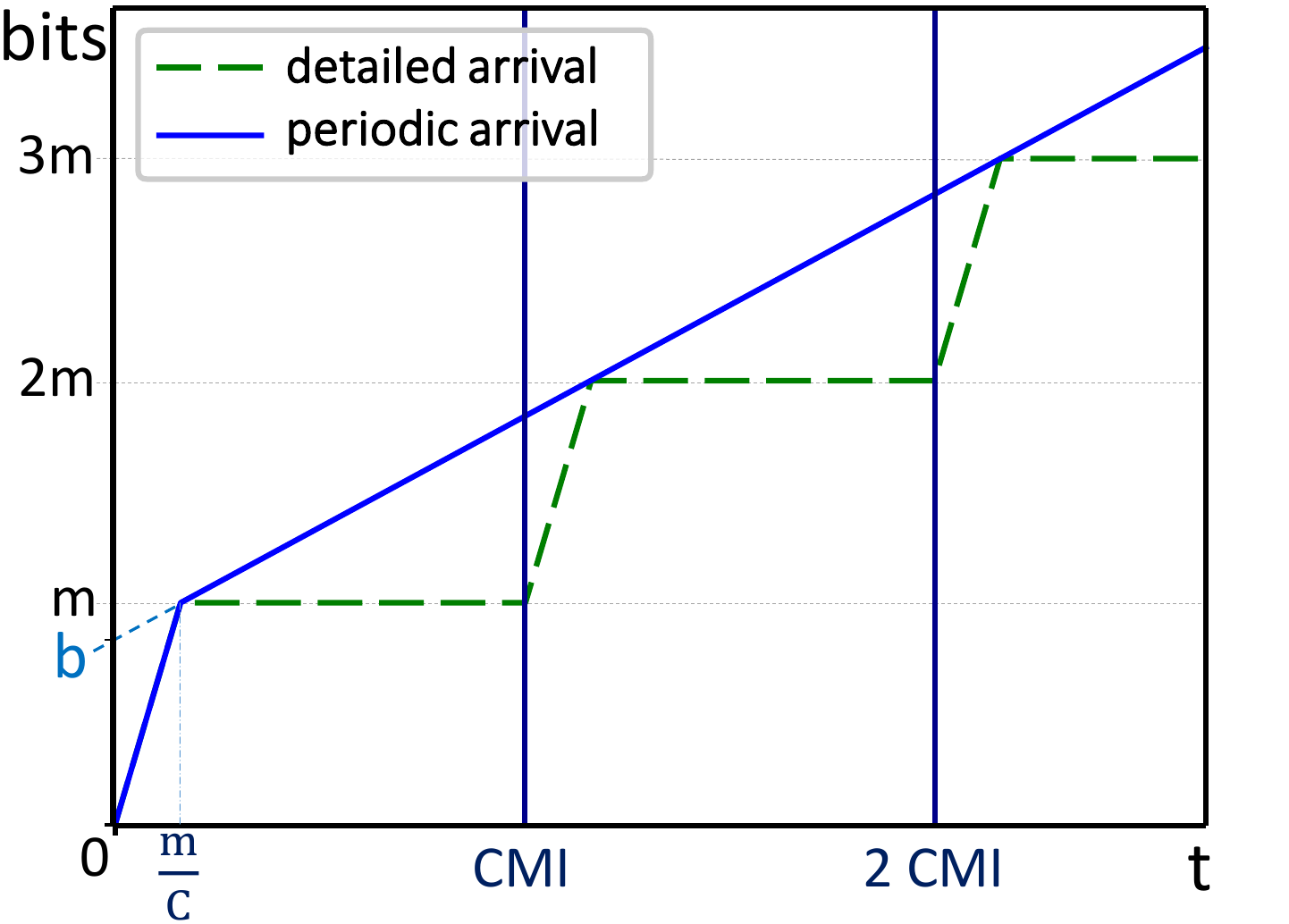} }}%
    \qquad
    \subfloat[Arrival curves for aperiodic traffic, with proposed correction.]{{\label{aper_arr}\includegraphics[width=4.9cm, page=2]{Bilder/pythoncurves/arrival.pdf} }}%
    \label{fig:ex_curves}%
    \caption{NC arrival curves.}
\end{figure*}

An arrival curve $\alpha(t)$ represents the maximum number of bits that arrive at the system during any period of length $t$. It is defined as
\cite{boudec_network_2012, borys_principles_2011, boudec_made_easy, boudec_I, BOUILLARD2010306}:
\begin {equation} R(t) \le (R\otimes \alpha)(t) \overset{def.}{=}  \inf_{0\le s \le t} \{R(t-s) + \alpha(s)\}\end {equation} 
In other words, during any time window of length $t$, the amount of additional data sent by the flow is limited
by $\alpha(t)$~\cite{kellerer_network_2016}.
The most commonly used arrival curve is called leaky-bucket (often also named rate-burst or token-bucket). It is defined as ${\alpha(t) = b + rt}$, where $b$ is the instantaneous burst and $r$ the constant rate.
Leaky-bucket curves have the advantage that the maximum delay and backlog can be obtained easily due to the simple nature of the curve as we will see in Section~\ref{sec:example}.

$\alpha(t)$ is derived by formal analysis of the flows, for example by measuring, defining flow requirements, or analyzing application code.
For TSN devices, the 802.1Qat standard~\cite{Qat} defines a characterization that a single flow can send at maximum MaxIntervalFrame (MIF) packets with MaxFrameSize (MFS) bytes in any interval of length ClassMeasurementInterval (CMI). A source might send periodically or aperiodically and the interval can either have fixed starting and ending times or is implemented as a sliding window~\cite{daigmorte_modelling_nodate}.

There are two ways to define arrival curves of sending TSN end-stations, either detailed, which models exactly the behavior as defined in the standard~\cite{Qat}, or simplified to ease calculations. The detailed curve was first introduced by De~Azua \textit{et~al.}~\cite{framework} and later discussed in detail by Daigmorte \textit{et~al.}~\cite{daigmorte_modelling_nodate}. Fig.~\ref{sending} shows the maximum output of sending TSN devices. The curve results from the assumption that a TSN end-node sends all data that it has reserved for CMI in the WC right at the beginning of an interval and with the maximum link rate $C$. This results in the step-like arrival curve as shown in Fig.~\ref{per_arr} by the green curve.
The formula for the detailed arrival curve for a periodic flow or any flow which is implemented with a sliding window is~\cite{framework, zhao_improving_2018, daigmorte_modelling_nodate}:
\begin {equation} \alpha'_{per}(t)=(m\cdot\lceil \frac{t}{CMI}\rceil ) \otimes (Ct) \end {equation} 
with $ m $ being the number of bits per interval ($ m = MIF\cdot MFS\cdot 8 $) and $C\cdot t$ modeling the maximum rate at which bits can be sent on the link.

The simplified leaky-bucket arrival curve upper bounds the detailed curve and is~\cite{framework, zhao_improving_2018, daigmorte_modelling_nodate}:
\begin {equation} \alpha(t)= \min(Ct,b+rt) \label{equ:simplearr}\end {equation} 
with $r = \frac{m}{CMI}$ and $b = m\cdot (1-\frac{r}{C})$. For an intuitive illustration, see the blue curve in Fig.~\ref{per_arr}.

If the sender is aperiodic, it might send twice the amount of bits right after another as seen in Fig.~\ref{sending} and~\ref{aper_arr}. The detailed arrival curve has therefore been denoted by De~Azua \textit{et~al.}~\cite{framework} and Daigmorte \textit{et~al.}~\cite{daigmorte_modelling_nodate} with:
\begin {equation} \alpha'_{per}(t)=(m\cdot\lceil \frac{t+CMI}{CMI}\rceil )
\label{equ:detailedarr} \otimes (Ct) \end {equation} 
However, as illustrated in red in Fig.~\ref{aper_arr}, this detailed arrival curve is greater than the described scenario and, therefore, also greater than the proposed simple arrival curve of the authors for aperiodic traffic. With the shaping effect of the link, the initial burst needs to be corrected to represent the desired behavior of the authors. Therefore, we propose the following detailed arrival curve as improvement instead of~\eqref{equ:detailedarr}:
\begin{multline}
\alpha'_{per}(t) = \sup_{0\le u\le t}\{C \cdot (u - \Bigl[ (CMI-\frac{m}{C})\cdot\lceil\frac{u-2\cdot\frac{m}{C}}{CMI}\rceil)\Bigr]^+ \}
\end{multline}
where $[x]^+$ means $\max(0,x)$. Thereby, the argumentation for this function is the same as before, but we adapted the formula to fit this reasoning. Illustratively, we send with the maximum link rate $C$ at the beginning of each CMI and then substract the time for which we do not send, which is $CMI-\frac{m}{C}$ for each interval. The current interval is calculated by $\lceil\frac{u-2\cdot\frac{m}{C}}{CMI}\rceil$. The supremum assures that the curve is non-decreasing.
 
We illustrate the simple arrival curve, the initial detailed arrival curve, and our improvement in Fig.~\ref{aper_arr}. As can be seen, the simple arrival curve formula is the same as for periodic traffic in~\eqref{equ:simplearr} but with a doubled initial burst~\cite{zhao_improving_2018, daigmorte_modelling_nodate}.

To account for the class-based scheduling, which means that all flows of the same class are scheduled in the same queue, the arrivals of all individual flows are summed up and are treated as aggregated input for the system~\cite{TT_CBS_CDT}.

\subsection{Service Curves for TSN}
\label{sec:service}
The details of packet scheduling in TSN are abstracted using service curves. $\beta(t)$ is the minimum service that a system offers to a flow during a period of length $ t $. $\beta(t)$ is defined as
\cite{boudec_network_2012, borys_principles_2011, boudec_made_easy, boudec_I, BOUILLARD2010306}:
\begin {equation} R'(t) \ge (R\otimes \beta)(t) \end {equation} 
A service curve $\beta$ thus guarantees that a system with input $R$ has at least $R \otimes \beta$ as output~\cite{kellerer_network_2016}. 
Service curves are the actual modeling part of NC, they are defined by a detailed analysis of the queuing, scheduling, and forwarding mechanisms. 
A commonly used form of service curve is the rate-latency function $\beta(t) = R\cdot[t-T]^+$, where $R$ is the minimum service rate after an initial WC delay of $T$.

A common notation for schedulers and their combinations helps to quickly identify the focus of published work in the future, as proposed by Luxi Zhao \textit{et~al.}~\cite{zhao_improving_2018} and Daigmorte \textit{et~al.}~\cite{daigmorte_modelling_nodate}. Contrary to these suggestions, we would not name best-effort traffic in the notation, since it is supposed to be always included in TSN with at least 25\%~\cite{802Q}. For the same reason, we would not mention SP in a notation for TSN schedulers. Thus, we propose a simple notation for the TSN schedulers according to the principle \textit{CBS, TAS}, or \textit{TAS-CBS} for scheduler combinations.

\subsubsection{CBS Service Curve}
\label{sec:CBS}
A frequently addressed scheduler in TSN is the Credit-Based Shaper (CBS). For CBS, each queue has a credit that decreases when traffic is sent and increases when the credit is negative or when scheduled traffic cannot be sent due to the transmission of other queues. If the queue is empty but the credit is positive, it is set to zero. As long as the credit of a queue is greater or equal to zero and no other packet is currently transmitted, a frame can be sent, otherwise, the packet is delayed. In contrast to the general assumption, not only class A and B can be implemented but an arbitrary number of CBS classes~\cite{Qav}. The rate with which the credit increases is called \textit{IdleSlope}, the decreasing rate \textit{SendSlope}.
CBS is defined in IEEE 802.1 Qav~\cite{Qav}.
For the subsequent analysis, we define the following variables:
\begin{itemize}
\item $V^x_{max}$: maximum CBS credit value for class $X$
\item $I^X$ and $S^X$: Idle- and SendSlope variables for class $X$
\item $l^X$: maximum packet size of class $ X $
\item $\bar{l}^X$: maximum packet size of all classes with lower priority than class $X$
\item $C$: maximum link rate
\end{itemize}

All NC publications that focus on the derivation of CBS curves result in the same service curve. To sum up all publications, the basic service curve for the CBS class X is a rate-latency curve and has the following form \cite{queck, framework, zhao_improving_2018}:
\begin {equation}\label{equ:cbsservice} \beta^X_{CBS}(t) = I^X \cdot (t-\frac{V^x_{max}}{I^X}) \end {equation} 
We rearranged this formula to show its intuitive behavior. Thereby, we assume that the slope parameters are defined with respect to the standard ($C = I^X-S^X~,~S^X < 0$ ). Since the IdleSlope is the allocated bandwidth for the flows~\cite{Qav}, it is also the rate parameter in this formula. $V^x_{max}$ models the highest credit that is possible for this queue, thus, flows of class $X$ are guaranteed to send after this WC value is reached, which is modeled as the latency part of \eqref{equ:cbsservice}.

\begin{figure}[!t]
\centering
\includegraphics[trim={0cm 1cm 0cm 0cm} ,clip,,width=5.4cm]{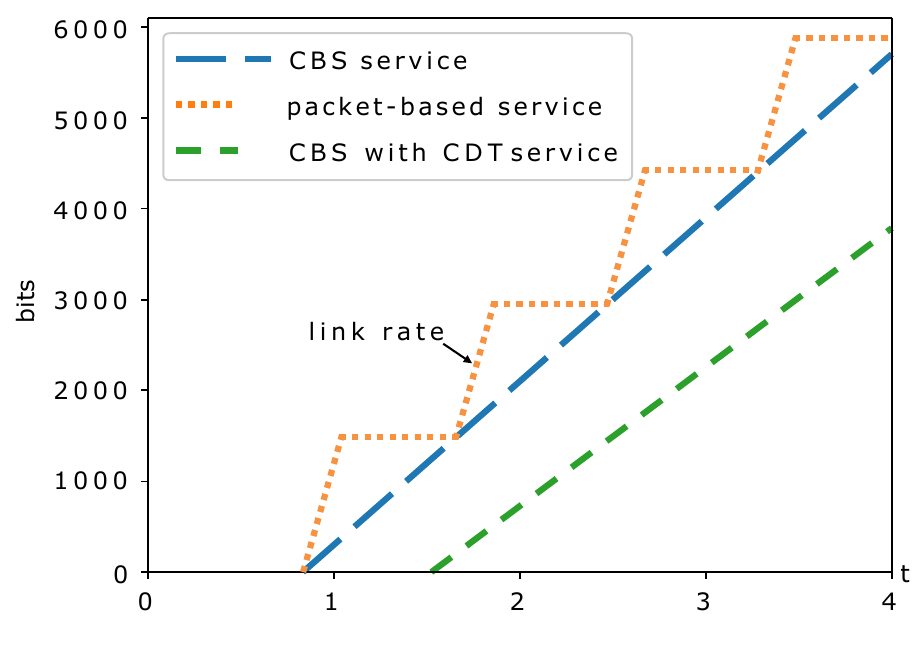}
\caption{CBS service curve with and without influence of CDT assuming example values and $t$ in seconds.}
\label{fig:improvements}
\end{figure}

The maximum credit $V^x_{max}$ has been improved several times, both for more classes and with tighter values. Reference~\cite{credit} present the existing credits and according to the authors derive tighter maximum bounds regarding an arbitrary number of classes. Thus, the maximum CBS credit of class $ X $ is~\cite{credit}:
\begin {equation} V^x_{max} = \frac{I^X}{C\cdot (C-\sum^{x-1}_{j=1}I^j)}\cdot (C\cdot \bar{l}^X-\sum^{x-1}_{j=1}S^j\cdot l^j) \end {equation}
For the highest CBS priority, often called class A, this equation simply leads to:
\begin {equation} V^1_{max} = \frac{I^1}{C}\cdot\bar{l}^1 \end {equation} 

Fig.~\ref{fig:improvements} shows the rate-latency behavior of the CBS service curve. As the individual packets under transmission are always sent with full outgoing link rate, the CBS curve can be improved as shown with the packet-based behavior in Fig.~\ref{fig:improvements}. This improves the delay bounds by 2-10\% at the cost of a more complex and packet-dependent curve definition, for details see~\cite{mohammadpour_known_transmission}.

\subsubsection{CBS Service Curve with CDT}
\label{sec:CBS_CDT}
Later, the idea of higher prioritized traffic than the CBS traffic frequently occurred in literature and is often referred to as Control Data Traffic (CDT)~\cite{alderisi_introducing_2013, boudec, zhao_improving_2018, CDT_Siemens}. If CDT is present, the initially defined CBS curve is adapted to model that the CBS queues cannot transmit during the transmission of CDT. Thus, the duration for which CDT is sent at time $ t $, named $\Delta t_{CDT}(t)$, is subtracted from the time parameter $ t $ of the CBS curve~\cite{boudec_short}:
\begin {equation} \label{cbs_cdt} \beta^X_{CBS\_CDT}(t) = \beta^X_{CBS}(t-\Delta t_{CDT}(t)) \end {equation} 
This clearly structured equation is one example for the contributions of this survey. With this clear notation, we see that the existing CBS service curve from \eqref{equ:cbsservice} can simply be reused with a small adaption on the time parameter.

We now discuss how the duration $\Delta t_{CDT}(t)$ for which CDT is sent can be defined. Since CDT only describes a higher prioritized queue than the CBS queue, the CDT arrival can be modeled as general leaky-bucket arrival curve $\alpha_{CDT}(t) = b + rt$ with the corresponding service curve $\beta_{CDT}(t) = C\cdot  [t-\frac{\bar{l}^{CDT}}{C}]^+$~\cite{boudec_short}. The service curve models that, in the WC, CDT needs to wait for a maximum of one other packet to finish transmission and can then send at full line speed.
Mohammadpour \textit{et~al.}~\cite{boudec_short} show that $\Delta t_{CDT}(t)$ is derived by dividing the output of the CDT $\Delta O(t)$ with the link rate $C$ (since generally: ${C\cdot\Delta t_{CDT}(t) = \Delta O(t)}$).  With a given arrival and service curve, NC defines a constraint on the maximum output $\alpha^*$ by using the min-plus deconvolution \cite{boudec_network_2012, boudec_made_easy, boudec_I}:
\begin {equation} a^*=\alpha \oslash \beta  \end{equation}
Thus, the output for CDT is derived with 
\begin {equation} \Delta O (t) \le (\alpha_{CDT} \oslash \beta_{CDT})(t)\end{equation}
As a result, $\Delta t_{CDT}(t)$ of \eqref{cbs_cdt} is upper bounded by~\cite{boudec_short}:
\begin {equation} \Delta t_{CDT}(t) \le \frac{(\alpha_{CDT} \oslash \beta_{CDT})(t)}{C} = \frac{b+r(t+ \frac{\bar{l}^{CDT}}{C})}{C} \end {equation} 
The CBS queues cannot send while CDT is sent, thus the overall service rate for the CBS service curve is decreased, which is also illustrated with the green curve in Fig.~\ref{fig:improvements}.

\subsubsection{TAS Service Curve}
\label{sec:TAS}
\begin{figure}[!t]
\centering
\includegraphics[trim={0cm 2.8cm 0cm 0cm} ,clip, page=2, width=2.4in]{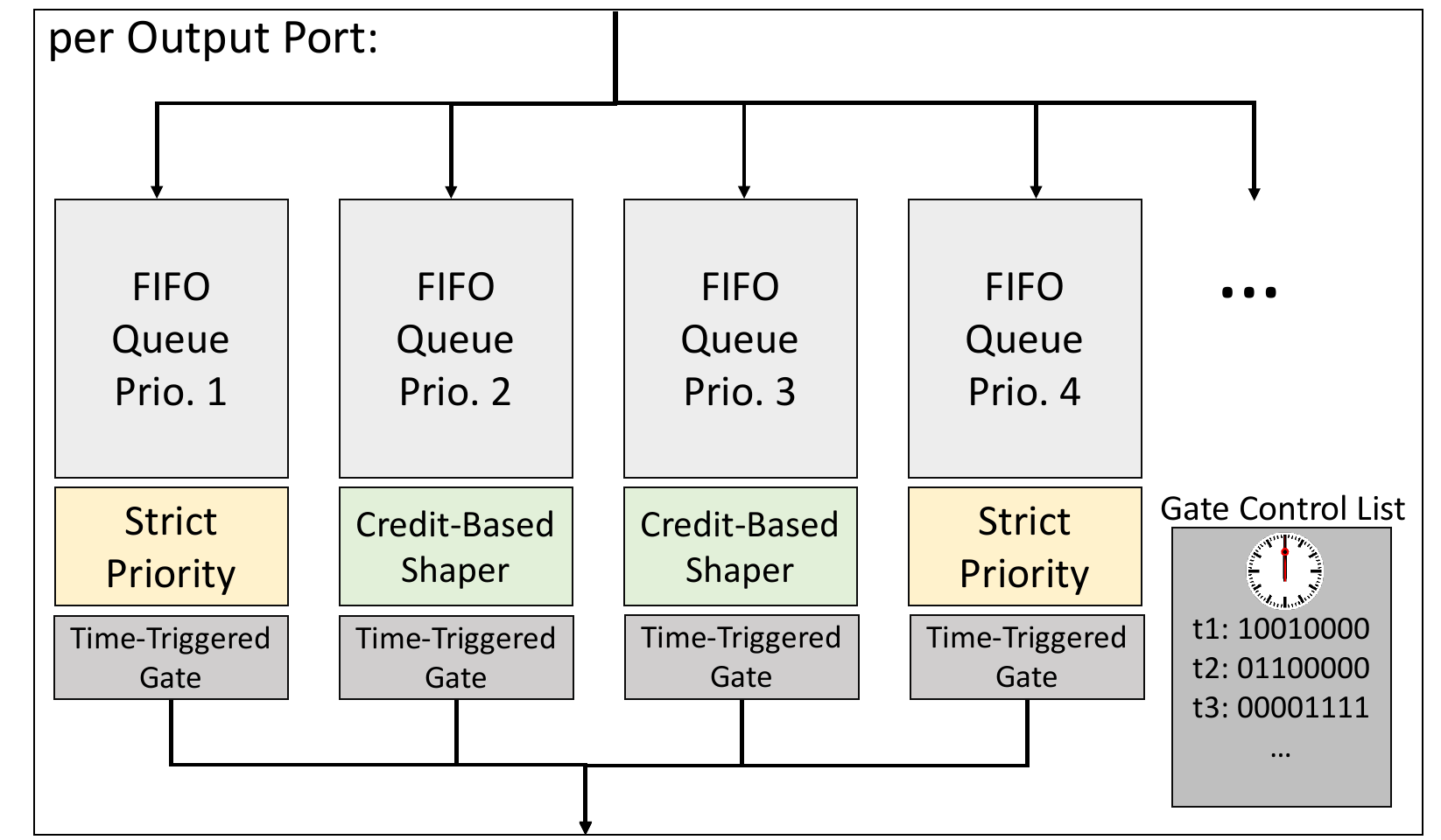}%
\caption{TAS scheduler with GCL on the right that defines which queues are open at which time.}
\label{Fig:TAS}
\end{figure}
Besides CBS, TSN introduced the Time-Aware Scheduler (TAS)~\cite{Qbv}. TAS works similar to TDMA but the time slots can be arbitrarily configured. Each queue has an associated gate and packets in queues will only be forwarded if the gate of this queue is open. The time slots for sending are called gate-open intervals. With Gate Control Lists (GCL), the gate-open intervals can be defined for each queue as illustrated in Fig.~\ref{Fig:TAS}. To ensure that the sending of a packet does not interfere with the gate-open intervals of another queue, guard bands are implemented which prevent the start of packet transmissions before critical gate-open intervals of other queues. The standard for TAS is IEEE 802.1Qbv~\cite{Qbv}. In addition, the TSN working group has defined the concept of frame preemption in the 802.1Qbu standard~\cite{Qbu} to interrupt the transmission of lower priority frames by higher priorities. This reduces the waste of bandwidth due to guard bands and offers less jitter and latency, which must also be considered for NC modeling.

Current approaches use TAS with strict separation conditions on the flows, meaning that only one gate is open and only packets of one flow are allowed in a queue at the same time~\cite{craciunas_scheduling_2016, anna}. Then the sent packet is repeatedly placed in an open queue at each node and can pass directly through the network. With these prerequisites, the delay calculation is completely deterministic and only technical delays need to be considered, but the possibility to buffer packets and also a large amount of bandwidth is lost. If we drop all separation conditions, the transmission no longer behaves deterministically and we are in the need of NC to derive guarantees. This has the great advantage that it offers more possibilities for reserving flows and greater flexibility in the configuration of the GCL~\cite{TT}.

\begin{figure}[!t]
\centering
\includegraphics[trim={0cm 5.5cm 0cm 0cm}, page=1 ,clip, width=2.4in]{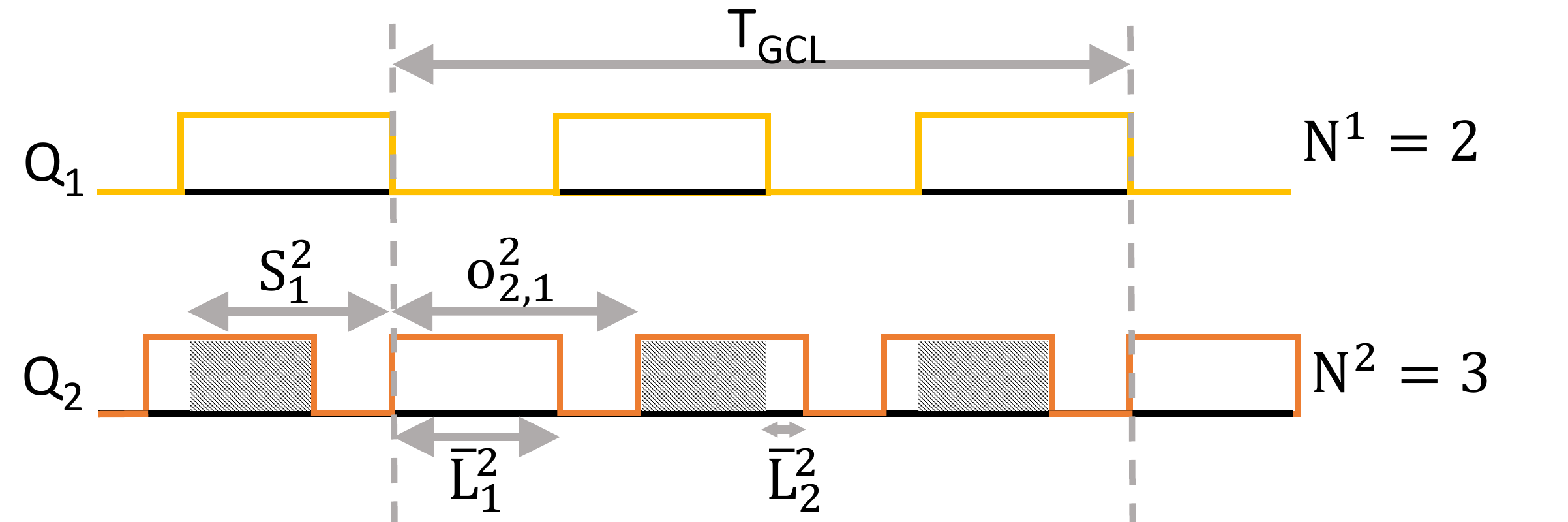}%
\caption{Illustration of variables for NC curves in TAS.}
\label{Fig:variables}
\end{figure}

Luxi Zhao \textit{et~al.}~\cite{TT} offer an NC model that can derive WC bounds for TAS without requiring any separation restrictions. 
Therefore, they we introduce additional variables that model the behavior of the GCL:
\begin{itemize}
\item $T_{GCL}$: hyperperiod of the GCL which is the least-common multiple of all GCL cycles for all classes
\item $\bar{L}^X_i$: guaranteed time slots during which a class $X$ has guaranteed service meaning that no higher priority gate-open intervals or guard bands overlap
\item $S_i^X$: the maximum waiting time between the beginning of a backlogged period and the timeslot $\bar{L}^X_i$ for class $X$
\item $o^X_{j,i}$: the relative offset which is the time between the starting of $\bar{L}^X_j$ and $\bar{L}^X_i$
\item $N^X$: the number of times that the gates of class $X$ open within $T_{GCL}$
\end{itemize}
All variables are visualized for two queues in Fig.~\ref{Fig:variables} where $Q_1$ has the highest priority. The curves illustrate the gate-open intervals. See Luxi Zhao \textit{et~al.}~\cite{TT} for a discussion on how to determine the guaranteed time slots by considering guard bands and non-preemptive packet transmission of other queues.

During a guaranteed time slot $\bar{L}^X_i$, a class can send with maximum link rate $C$ since not other traffic can interfere. As a result, the overall TAS service curve for class $X$ can be derived by only analyzing the GCL. It is defined by the following two functions~\cite{TT}:
\begin {equation}\label{equ:sum_TAS} \beta^X_{TAS}(t) = \sum^{j+N^X-1}_{j=i} \beta_{T_{GCL},\bar{L}_j^X}(t+T_{GCL}-\bar{L}_j^X-S_i^X-o^X_{j,i}) \end {equation} 
which sums up functions of the form $\beta_{T,L}$ as~\cite{TDMA}:
\begin {equation}\label{equ:single_TAS} \beta_{T,L}(t) = C\cdot\max(\lfloor \frac{t}{T} \rfloor \cdot L, t-\lceil \frac{t}{T}\rceil \cdot(T-L)) \end {equation}
With these functions, we model the latest appearance for each guaranteed time slot separately. Equation~\eqref{equ:single_TAS} is the normal TDMA service curve as derived in~\cite{TDMA} for one gate-open interval. Then, equation~\eqref{equ:sum_TAS} is the overall service curve which sums up the curves for all guaranteed time slots of class $X$. 
In other words, $\beta^X_{TAS}(t)$ models the latest possible time at which the guaranteed time slots can appear and that only during this time, class $X$ can send with rate $C$.

The resulting service curve is shown in Fig.~\ref{Fig:TAS-CBS}, the left side of~\eqref{equ:single_TAS} is highlighted in green, the right side in blue and the resulting curve is the maximum of both. 

\subsubsection{CBS Service Curve with TAS}
\label{sec:TAS_CBS}
We now model the service curve for CBS queues that are used in combination with TAS. 
If combined with TAS, the CBS credit is frozen if a CBS gate is closed. Besides, all existing NC approaches assume that the credit is frozen if transmission is forbidden, for example, due to guard bands. A discussion about the credit evolution in TAS-CBS can be found in~\cite{boyer_credit_2019}. The gates are commonly configured in a way that all CBS and BE queues are open at the same time when all purely time-triggered gates are closed and CDT is simply implemented as a higher prioritized time-triggered queue~\cite{TT_CBS_CDT, zhao_improving_2018, zhou_analysis_2018, daigmorte_modelling_nodate}.
Therefore, CDT does not need to be regarded separately to other time-triggered queues.

Using the previously defined service curve $\beta^X_{CBS}$, we can derive the adapted TAS-CBS service curve by subtracting the time that the gates of the CBS queues are closed. The approach is similar to $\beta^X_{TAS}$ that we derived before but instead of modeling the minimum time that the CBS queues are open, we now subtract the maximum time that the other gates are open for an easier calculation. As stated above, we assume that all CBS queues are open if the other time-triggered queues are closed. For this, we need an additional variable $L^{TT}_i$ which denotes a complete gate-open interval for all non-CBS queues, not just the guaranteed time slots. The maximum time that all non-CBS queues are open can be modeled by~\cite{zhao_improving_2018, daigmorte_modelling_nodate}:
\begin {equation}\label{equ:CBSTAS} f(t) = \max_{0 \le i \le N^{TT}-1}\{ \sum^{j+N^{TT}-1}_{j=i} L^{TT}_j \cdot \lceil \frac{t-o^{TT}_{j,i}}{T_{GCL}}\rceil \} \end {equation} 
Thereby, $\lceil \frac{t-o^{TT}_{j,i}}{T_{GCL}}\rceil$ is the earliest time that $ L^{TT}_j$ can occur with reference to $L^{TT}_i$. The maximum in \eqref{equ:CBSTAS} ensures that the reference index $i$ is optimized. Similar to the effect of CDT, the TAS-CBS service curves is then simply derived by subtracting the time during which other queues are open from the CBS service curve:
\begin {equation}\label{equ:tas_cbs} \beta^X_{TAS\_CBS}(t) = \beta^X_{CBS}(\sup_{0\le u \le t}\{u-f(u)\})\end {equation} 

\begin{figure}[!t]
\centering
\includegraphics[width=5.4cm]{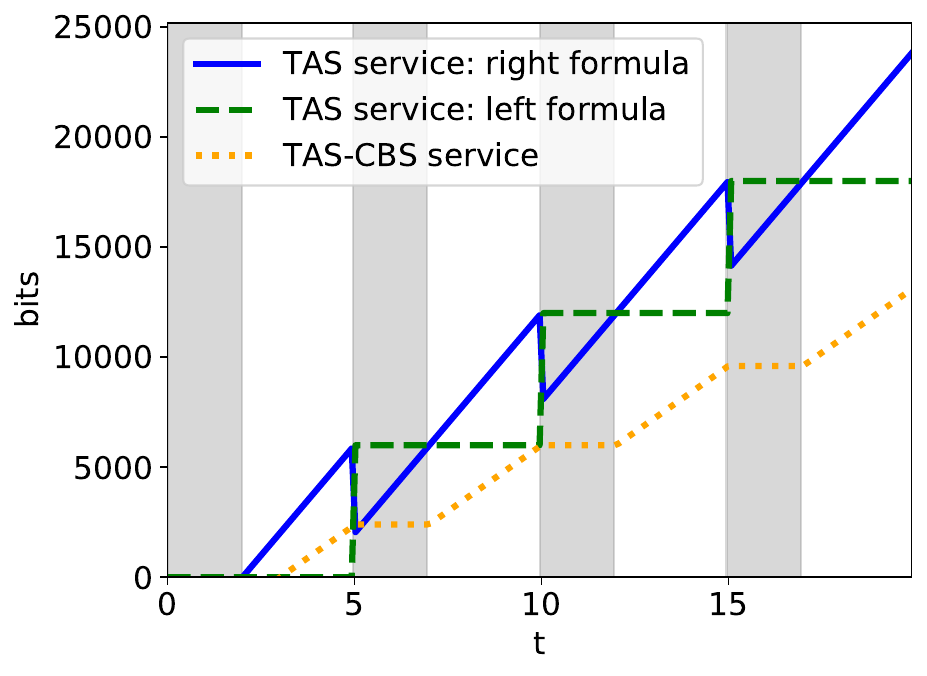}%
\caption{TAS and TAS-CBS service curves assuming example values and $t$ in seconds. White areas are guaranteed time slots. Be aware that TAS gates can send with full link rate whereas CBS gates can only send with their reserved bandwidth $I^X$.}
\label{Fig:TAS-CBS}
\end{figure}

 \begin{figure*}%
     \label{fig:allExample}%
     \begin{minipage}{.55\textwidth} 
    \subfloat[ATS scheduler with interleaved regulators that privilege higher prioritized traffic.]{{\label{Fig:ATS}\includegraphics[trim={0cm 2.8cm 0cm 0cm} ,clip, page=3, width=2.1in]{Bilder/TSNprinciples.pdf} }}%
        \qquad
    \subfloat[Delay analysis for an ATS-CBS scheduler.]{{\label{ATS-CBS}\includegraphics[trim={0cm 0cm 0cm 0cm}, clip, width=1.2in]{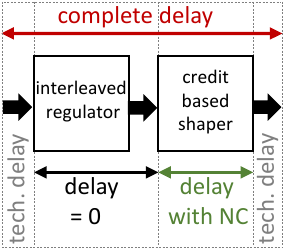} }}%
            \caption{Asynchronous Traffic Shaping with Credit Based Shaper.}  
     \end{minipage}%
          \begin{minipage}{.45\textwidth} 
                    \subfloat[Arrival $\alpha$, service $\beta$, and output $\alpha^*$ curves with  $\alpha^* = \alpha \oslash \beta$.]{{\label{NCcurves_graph}\includegraphics[width=1.5in, page=1]{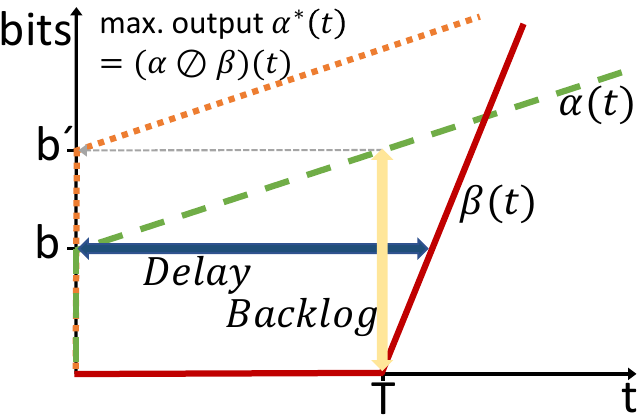} }}%
        \qquad
    \subfloat[Small example network.]{{\label{fig:example_net}\includegraphics[width=1.3in]{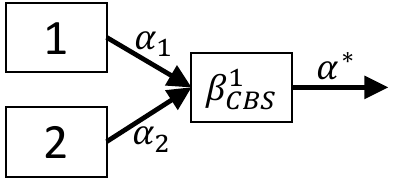} }}%
        \caption{Examples for Network Calculus principles.}
             \end{minipage}%
  \end{figure*}

The supremum in~\eqref{equ:tas_cbs} ensures that the service curve is non-decreasing. Fig.~\ref{Fig:TAS-CBS} shows $\beta^X_{TAS\_CBS}(t)$ in orange. In fact, equation \eqref{equ:CBSTAS} needs to be refined to include the effect of guard bands and possibly preemption. Therefore, Luxi Zhao \textit{et~al.}~\cite{zhao_improving_2018} introduce additional variables to represent guard bands and adapt \eqref{equ:CBSTAS} accordingly. We do not present these variables and adaptions to improve the comprehensibility as \eqref{equ:CBSTAS} already shows the basic behavior. For the service curve with guard bands, see (10) in~\cite{zhao_improving_2018}. Daigmorte \textit{et~al.} define the same behavior but with a different notation, for details see (12) in~\cite{daigmorte_modelling_nodate}. NC models for a different evolution of the CBS credit have not been derived yet.

For the preemption mode, we have different approaches in~\cite{daigmorte_modelling_nodate} and~\cite{TT_CBS_CDT}. They have a common successor paper~\cite{zhao_improving_2018} where preemption is no longer discussed. Therefore, we want to emphasize that the modeling of preemption in TSN with NC is not clear yet. The difference is the treatment of overheads (headers and/or trailers) that are added to the preempted frames to identify and map the corresponding parts of the packets. Reference~\cite{TT_CBS_CDT} models an increase of the credit during the transmission of overheads, while~\cite{daigmorte_modelling_nodate} models that the credit is frozen.
An analysis of different implementations for the credit behavior is still left for future research.
    
\subsubsection{ATS}
\label{sec:ATS}

Instead of gates, the TSN scheduling mechanism Asynchronous Traffic Shaping (ATS) uses interleaved regulators to shape incoming traffic. As seen in Fig.~\ref{Fig:ATS}, the packets are first stored in the interleaved regulators and then forwarded to the output queues, thereby, flows are reshaped according to their priorities.
ATS has rules to separate traffic in the interleaved regulators, so higher-priority packets can pass low-priority packets without the need for dedicated per-flow queues~\cite{specht_urgency-based_2016, specht_synthesis_2017}.
A major advantage of this scheduler is that traffic is reshaped at each hop so bursts of flows are limited. Therefore, the burstiness of flows does not successively increase. 
The effect of interleaved regulators has been analyzed with NC by Le Boudec~\cite{ir}. The result is that ATS does not add any end-to-end delay to already shaped sources. Thus, they potentially increase the delay only at the first hop. 

\subsubsection{CBS Service Curve with ATS}
\label{sec:ATS_CBS}
Mohammadpour \textit{et~al.}~\cite{boudec_short} published the only NC work dealing with ATS-CBS curves. Instead of adapting the service curve, they model the scheduler combinations as series-connected network elements as shown in Fig.~\ref{ATS-CBS}. Thus, the service curve remains unchanged to the previously introduced curves $\beta^X_{CBS}$ or $\beta^X_{CBS\_CDT}$. Then they use the NC ruleset to derive the maximum delay for the CBS element and additionally add technical delays of the network. Besides, they use the result of~\cite{ir} which proves that minimal interleaved regulators as implemented in ATS do not increase the WC delay if flows are shaped at the sending end-devices. Thus, Mohammadpour \textit{et~al.}~\cite{boudec_short} show that the NC curves for ATS-CBS are the same as the previously presented curves for CBS.

\subsection{Derived Guarantees}
\label{sec:guarantees}

Only by defining the arrival curve $\alpha(t)$ and a service curve $\beta(t)$, NC can provide three guaranteed WC bounds. The three guarantees are: delay, also called latency; backlog, also called memory or buffer size; and maximum output.

As illustrated in Fig.~\ref{NCcurves_graph}, the WC delay is simply the maximum horizontal distance between the arrival and service curve of a flow. It is formally defined as
\cite{boudec_network_2012, boudec_made_easy, boudec_I, BOUILLARD2010306}:
\begin {equation} D \le \sup_{0\le s}\{\inf \{\tau \ge 0 \mid \alpha (s) \le \beta (s+\tau )\}\}\end{equation}

The WC backlog is the number of bits that are currently in the system, so it is defined as the maximum vertical distance between the arrival and service curve
\cite{boudec_network_2012, boudec_made_easy, boudec_I, BOUILLARD2010306}:
\begin {equation} B \le \sup_{0\le s}\{\alpha (s) - \beta (s) \}\end{equation}

\subsection{Example for CBS}
\label{sec:example}

\begin{figure}[!t]
\centering
\includegraphics[width=0.5\textwidth]{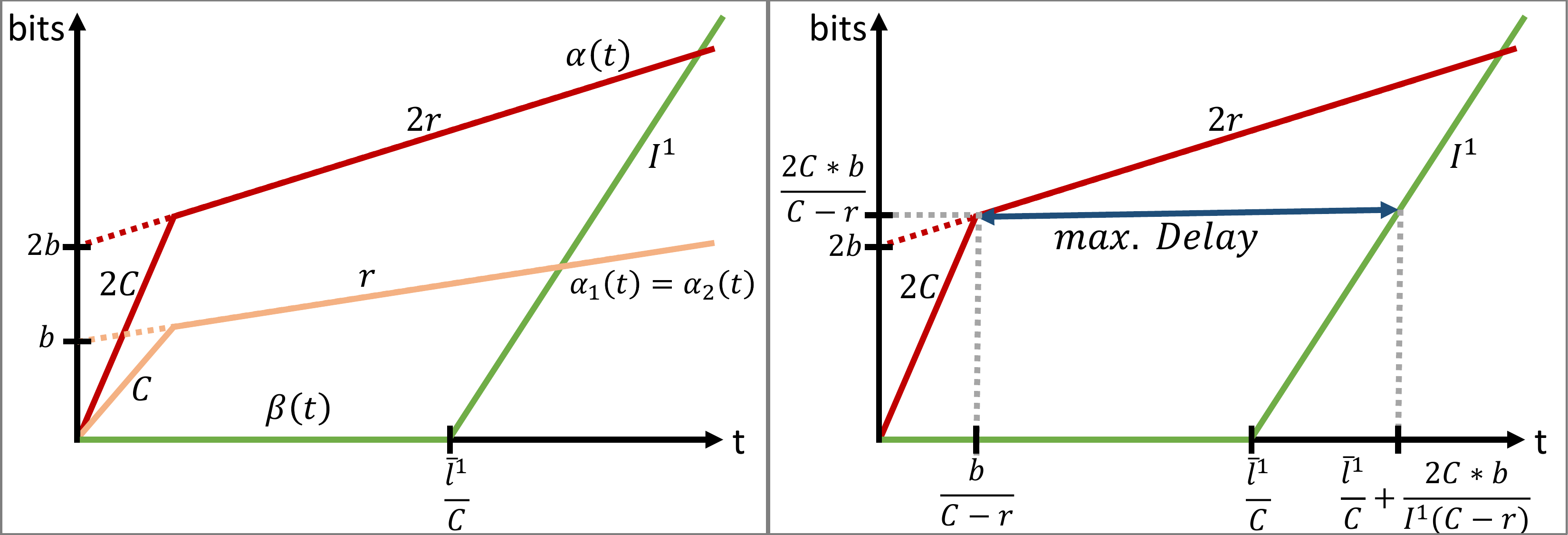}
\caption{Delay calculation for an example network.}
\label{fig:example_calc}
\end{figure}

To demonstrate how NC derives delay guarantees, assume the small one-hop network in Fig.~\ref{fig:example_net}. In this scenario, we have two sources each sending one flow of the highest priority that passes a forwarding device that schedules traffic using CBS. Without loss of generality, we assume that the two flows of the sources are identical. We want the WC queuing and scheduling delay that can occur at this forwarding device. 
\begin{table*}[!h!t]
\renewcommand{\arraystretch}{1.3}
\centering
\caption{A short overview of existing NC work. All mechanisms are listed.}
\label{table1}
\resizebox{\textwidth}{!}{%
\begin{tabular}{|c|c|c|c|c|c|c|c|c|c|c|c|c|c|} 
\hline
Source & Mechanism                               & Author                                                                                                 & Year       & \begin{tabular}[c]{@{}c@{}}Pre-\\ emption \end{tabular} & Work Basis & \begin{tabular}[c]{@{}c@{}}Impact\\ of CDT \end{tabular} & \begin{tabular}[c]{@{}c@{}}Arrival\\ $\alpha$ \end{tabular}  & \begin{tabular}[c]{@{}c@{}}Min. Service\\ $\beta$ \end{tabular}                                                & \begin{tabular}[c]{@{}c@{}}Max. Service\\ $\beta^{max}$ \end{tabular} & \begin{tabular}[c]{@{}c@{}}Shaper\\ $\sigma$ \end{tabular}  & \begin{tabular}[c]{@{}c@{}}Max. Output\\ $\alpha^*$\end{tabular}                & Delay  & Backlog  \\ 
\hhline{|=|=|=|=|=|=|=|=|=|=|=|=|=|=|} 
       \cite{queck}  & CBS                                     & R. Queck                                                                                               & July 2012  & -                                                                               &   \cite{georges_strict_2005}           & no                                                                               & leaky-bucket      & \begin{tabular}[c]{@{}c@{}}min. 2 CBS\\  \& x SP \end{tabular}                                                &                                                                                               &                  & CBS \& SP                                                                                        & CBS \& SP &          \\ 
\hline
   \cite{framework}     & {\cellcolor[rgb]{1,0.502,0.251}}CBS     & \begin{tabular}[c]{@{}c@{}}J. A. Ruiz De \\ Azua \textit{et~al.} \end{tabular} & Oct. 2014  & -                                                                               &           \cite{queck} & no                                                                               & detailed          & \begin{tabular}[c]{@{}c@{}}min. \& strict \\ for 2 CBS \\ \& strict for SP \end{tabular} & 2 CBS                                                                                         & 2 CBS            &                                                                                                         &        &          \\ 
\hline
       \cite{zhao_improving_avb_2018} & CBS                                     & Lin Zhao \textit{et~al.}                                                                                & Nov. 2018  & -                                                                               &            \cite{framework}  & no                                                                               & detailed          & min. 3 CBS                                                                                                                         &                                                                                               &                  &                                                                                                         &        &          \\ 
\hline
       \cite{TT} & {\cellcolor[rgb]{1,0.502,0.251}}TAS     & Luxi Zhao \textit{et~al.}                                                                                & July 2018  & no                                                                              &          \cite{TT_CBS_CDT} \cite{TDMA}  & -                                                                                & leaky-bucket      & min. TT                                                                                                                            &                                                                                               &                  & TT & TT     &          \\ 
\hline
       \cite{flowshaping} & TAS-CBS                                 & F. He \textit{et~al.}                                                                                  & May 2017   & no                                                                              &           \cite{framework} & yes                                                                              & leaky-bucket      &                                                                                                                                        &                                                                                               & 2 CBS            & \begin{tabular}[c]{@{}c@{}}CBS incl. \\ shaper\end{tabular}                 &        &          \\ 
\hline
       \cite{TT_CBS_CDT} 	& TAS-CBS                                 & Luxi Zhao \textit{et~al.}                                                                                & April 2018 & yes\&no                                                                           &           \cite{framework}  & yes                                                                              & leaky-bucket      & min. 2 CBS                                                                                                                         &                                                                                               &                  & CBS                                                                                          & CBS    &          \\ 
\hline
      \cite{daigmorte_modelling_nodate} & TAS-CBS                                 & \begin{tabular}[c]{@{}c@{}}H. Daigmorte\\ \textit{et~al.} \end{tabular}        & June 2018  & yes\&no                                                                             &           \cite{TT_CBS_CDT} \cite{framework} & yes                                                                              & detailed          & \begin{tabular}[c]{@{}c@{}}min. \& strict\\ for x CBS \end{tabular}                                           &                                                                                               & x CBS            &                                                                                                         &        &          \\ 
\hline
       \cite{zhao_improving_2018} & {\cellcolor[rgb]{1,0.502,0.251}}TAS-CBS & Luxi Zhao \textit{et~al.}                                                                                & Dec. 2018  & no                                                                           &           \cite{daigmorte_modelling_nodate} \cite{TT_CBS_CDT}	\cite{framework}  & yes                                                                              & leaky-bucket      & min. x CBS                                                                                                                         &                                                                                               & x CBS \& link       & \begin{tabular}[c]{@{}c@{}}CBS incl.\\ shaper\end{tabular} & CBS    &          \\ 
\hline
      \cite{boudec_short} \& \cite{boudec} & {\cellcolor[rgb]{1,0.502,0.251}}ATS-CBS & \begin{tabular}[c]{@{}c@{}}E. Mohammad-\\ pour \textit{et~al.} \end{tabular}   & Sep. 2018  & -                                                                               &            \cite{framework}& yes                                                                              & =$\alpha^*$       & \begin{tabular}[c]{@{}c@{}}min. 2 CBS /\\ min. x CBS\cite{credit} \end{tabular}                                                                                                                         &                                                                                               &                  & \begin{tabular}[c]{@{}c@{}}CBS incl. \\ link \end{tabular}                & CBS  & CBS    \\
\hline
\end{tabular}
}
\end{table*}
The service curve for the CBS node is:
\begin {equation} \beta^1_{CBS}(t) = I^1 \cdot (t-\frac{\bar{l}^1}{C}) \end {equation} 
as explained in Section~\ref{sec:CBS}.
For the complete system, we need the aggregated arrival curve $\alpha(t)$ of all input sources which is $\alpha(t)=\alpha_1(t)+\alpha_2(t) = 2\cdot \min(Ct,b+rt)$.
How to derive the values for $b$ and $r$ is explained in Section~\ref{sec:arrival}.
The result is illustrated in Fig.~\ref{fig:example_calc} which shows the aggregation of the arrival curves on the left side and the delay calculations on the right side. As can be seen, the maximum queuing and scheduling delay can be derived as the maximum horizontal deviation between the arrival and service curve:
\begin {equation} D \le \frac{\bar{l}^1}{C} + \frac{2C\cdot b}{I^1(C-r)} - \frac{b}{C-r}  \end {equation} 
\subsection{Concatenation of Nodes}
\label{sec:concatenation}
Using NC, the delay for a single flow crossing a sequence of servers can be computed tightly, but if several flows are aggregated and scheduled as one class, as in TSN, computing tight end-to-end delays for flows over multiple hops can become very difficult. Note that all publications that are not using ATS offer pessimistic delays due to the class-based curves and the burstiness increase at each hop. Defining tight bounds for complex networks is still an ongoing research topic.
Fidler~\cite{fidler} provides a survey on the current state of the art for which network topologies tight bounds have been derived.

\subsection{Details of NC publications}
\label{sec:details}

The previous section presented arrival and service curves and illustrated how they use the results of each other. We now present the scope of each work that derived NC results for TSN. Besides arrival and minimum service curves, NC also models strict minimum service~\cite{BOUILLARD2010306}, maximum service~\cite{boudec_network_2012, borys_principles_2011}, and shaper curves~\cite{boudec_network_2012, borys_principles_2011, boudec_made_easy, boudec_I} which can improve the maximum output of systems. For the definitions, see~\cite{boudec_network_2012}.

Table~\ref{table1} is a compact overview of all papers that derive NC results for TSN. 
To the best of our knowledge, this is the complete list of all literature dealing with the derivation of NC curves for TSN schedulers. The table demonstrates how the papers build on each other in the column "Work Basis" in Table~\ref{table1}. The orange color indicates that the papers contain the most recent or advanced NC results for the corresponding mechanisms and can be the first source for lookup. Mohammadpour \textit{et~al.} have two similar papers, a published version~\cite{boudec_short} and an unpublished~\cite{boudec} in which proofs are included. As we can see, ATS is not in the table since it was only analyzed in general, not specifically for TSN.

Column "Preemption" lists whether the scheduler has been modeled with or without a preemptive behavior (indicated by "yes" or "no") or whether preemption is not relevant for this mechanism (indicated by "-"). Using the same notation, "Impact of CDT" illustrates if the work also models the effect of higher priority traffic than handled in the CBS queues. Column "Arrival" shows if they use the detailed or a simplified leaky-bucket assumption for the input modeling. However, with the results of De~Azua \textit{et~al.}~\cite{framework} and Daigmorte \textit{et~al.}~\cite{daigmorte_modelling_nodate}, generally applicable and standard conform values for leaky-bucket curves have been derived, as explained in detail in Section~\ref{sec:arrival}. Mohammadpour \textit{et~al.}~\cite{boudec_short} assume the arrival curve to be already shaped and thus being the same as the output of each hop which is denoted as "$=\alpha^*$".
The columns "Min. Service" to "Shaper" denote which curves were defined by the authors for which scheduling mechanism and for how many classes of the same mechanism. For example, "2 CBS" denotes that CBS curves for two priorities were derived whereas "x" symbolizes an arbitrary number of queues. 

Specifically, the column "Shaper" indicates whether the authors derived shaper curves and whether they considered the shaping effect of the link. The next column "Max. Arrival" then presents whether they provide a formula for the output and whether they used their shaper curve or the shaping of the link to improve this output. 

Finally, the last two columns show which guarantees were explicitly analyzed in the paper. However, with $\alpha$ and $\beta$ defined, the results for the delay, backlog, and maximum output flow can be derived even without a dedicated discussion as seen in the previous section. 

In the following, we present details on the goals, contributions, and problems of the individual papers of Table~\ref{table1}.

\subsubsection{R. Queck \cite{queck}}
With his work published in 2012, R.~Queck was the first to derive NC limits for CBS. He uses the results for strict priority (SP) and weighted fair queuing from Georges \textit{et~al.}~\cite{georges_strict_2005}. He derived service curves for two CBS and an arbitrary number of SP classes and examines the burstiness increase after each hop. Queck also highlighted that the delays need to be calculated successively due to this increase of burstiness.
The derived CBS service curve is still state of the art, as described in the previous section. However, the proof for the curve is more formal in De~Azua \textit{et~al.}~\cite{framework}.

\subsubsection{De~Azua \textit{et~al.} \cite{framework}}
De~Azua \textit{et~al.} used the work of R. Queck~\cite{queck} and extended it with the detailed arrival curve, additional service and shaper curves, and more formal proofs. In addition to Queck's minimum service curve, they provide strict minimum, maximum, and shaper curves for two CBS classes. They also offer a service curve for lower prioritized SP traffic, but not further divided into different SP classes. Shaper curves for CBS are typical for their publications in TSN. However, the authors later utter doubt about their proof for the maximum service curve~\cite{daigmorte_modelling_nodate}.

\subsubsection{Lin Zhao \textit{et~al.} \cite{zhao_improving_avb_2018}}
This work enhances the work of De~Azua \textit{et~al.}~\cite{framework} with a service curve for a third CBS class by bounding its credit.

\subsubsection{Luxi Zhao \textit{et~al.} \cite{TT}}
Luxi Zhao \textit{et~al.} have the only work that considers curves in TAS by analyzing the GCL without restrictions on the flow separation. However, they only assume the non-preemption mode which could be a possible task of improvement for the future.

\subsubsection{He \textit{et~al.} \cite{flowshaping}}
He \textit{et~al.} investigate the shaping benefit of CBS. Thus, they model shaper curves for two CBS classes with the influence of TAS traffic. Thereby, they investigate the optimal Send- and IdleSlope values. However, NC is only a small part of their analysis and the work is not as relevant for the NC analysis as the others.

\subsubsection{Luxi Zhao \textit{et~al.} \cite{TT_CBS_CDT}}
This is the first in a series of papers by Luxi Zhao \textit{et~al.} dealing with TAS-CBS. In this paper, they derive the detailed TAS-CBS service curve with which they obtain the delay for two CBS classes for both preemption and non-preemption.

\subsubsection{Daigmorte \textit{et~al.} \cite{daigmorte_modelling_nodate}}

The work of Luxi Zhao \textit{et~al.}~\cite{TT_CBS_CDT} and this paper are quite similar, written by the same authors on the same mechanism. This paper is an internal report with more details and the improved arrival curve of De~Azua \textit{et~al.}~\cite{framework}. As mentioned before, they model the gate-open intervals for CBS queues and the credit behavior for preemption overheads differently.

\subsubsection{Luxi Zhao \textit{et~al.} \cite{zhao_improving_2018}}

This is the improved work for Luxi Zhao \textit{et~al.}~\cite{TT_CBS_CDT}. They derive curves for an arbitrary number of CBS classes and slightly improve their results by introducing shaper curves for the link and the CBS in the maximum output curve.

\subsubsection{Mohammadpour \textit{et~al.} \cite{boudec_short}}
This is a comprehensive analysis of all delays in an ATS-CBS network. They are the first to analyze the impact of CDT for non-TT mechanisms. They use the work of~\cite{ir} and analyze the complete end-to-end delay since the reshaping allows an independent calculation of the delays at each hop. Mohammadpour \textit{et~al.} also improved the CBS service curve later in~\cite{credit} by deriving upper credit bounds for arbitrary numbers of CBS classes and thereby updated their service curves.

\section{Summary}
\label{sec:summary}

Network Calculus has various applications in TSN and the work still continues. We have provided an introduction to all the necessary information for using NC for TSN schedulers.

Reasons for NC in TSN are to guarantee upper limits for non-deterministic scheduler, gain information about the behavior of the schedulers and thus make design decisions, derive reliable evidence for upper network limits for critical applications, test the validity of TSN configurations, increase the solutions space for GCL, eliminate packet loss due to buffer overflows, automate the analysis with implementations for example in the Centralized Network Controller, and update the accumulated latency field in decentralized reservation protocols for TSN networks. 

Future work should investigate which scheduler or scheduler combination is best suited for which scenario or setup. With the overview in Table~\ref{table1}, we showed that all relevant TSN scheduler and combinations have been investigated, but some curves are still missing. Especially the credit evolution and preemption mode for TAS and TAS with CBS has not been modeled conclusively so far. 
Besides, it would be particularly important to investigate the increase of burstiness without reshaping of flows for complex and large networks and thereby analyze the quality of end-to-end delays.

To the best of our knowledge, we have mentioned every relevant literature related to TSN and NC. We proposed an improvement for the arrival curve of aperiodic TSN traffic while offering a uniform representation and a detailed description of all existing results that deal with NC for TSN. We are the first to discuss how the results of different publications built on or differ from each other. Finally, we provided brief summaries and a clear mapping of the scope for each paper and thus facilitate access to these topics for the interested audience.

\bibliographystyle{IEEEtran}
\bibliography{IEEEabrv,TSN_u_NC}

\end{document}